\documentclass[doublecol,linenumbers]{epl2} 
\usepackage{dcolumn}
\usepackage{graphicx}
\usepackage{amsmath}
\usepackage{amsfonts}
\usepackage{amssymb}

\title{Dynamical probe of thermodynamical properties in three-dimensional hairy AdS black holes}
\shorttitle{Title} 

\author{De-Cheng Zou\inst{1} \and Yunqi Liu\inst{2} \and Cheng-Yong Zhang\inst{3}
\and Bin Wang\inst{4}}
\shortauthor{De-Cheng Zou \etal}

\institute{
  \inst{1} College of Physical Science and Technology, Yangzhou University - Yangzhou, 225009, China\\
  \inst{2} School of Physics, Huazhong University of Science and Technology - Wuhan, 430074, China\\
  \inst{3} Center of High Energy Physics, Peking University - Beijing, 100871, China\\
  \inst{4} Department of Physics and Astronomy, Shanghai Jiao Tong University - Shanghai 200240, China
}

\abstract{
We study separatively the quasinormal modes (QNM) of electromagnetic perturbations
around three-dimensional anti-de Sitter(AdS) black holes in Jordan and Einstein frames,
which are related by the conformal transformations and a redefinition of a scalar field.
We find that, in the Jordan frame, the imaginary parts of QNM
frequencies can reflect the thermodynamical stabilities of hairy black holes, including
the possible phase transition between the hairy black hole and BTZ black hole,
disclosed by examining the corresponding free energies.
Similar results are also uncovered in the Einstein frame.
The obtained results further support that the QNM can be
a dynamic probe of the thermodynamic properties in black holes.}

\begin{document}

\maketitle

\section{Introduction}

In black hole physics there is a famous no hair
theorem. In general, the no-hair theorem rules
out the black holes coupled with a
scalar field in asymptotically flat
spacetimes\cite{Bekenstein:1974sf, Xanthopoulos:1992fm,Klimcik:1993cia},
because the scalar field can diverge on the
horizon and make the black hole become unstable\cite{Bronnikov:1978mx}. The dawn of
constructing a black hole with scalar hair broke
on the horizon when a negative cosmological
constant was taken into account. For example, AdS black holes with scalar
hair in the scalar-tensor theories (STT) have been respectively constructed
in three dimensional\cite{Martinez:1996gn,Henneaux:2002wm,Xu:2013nia,
Zhao:2013isa,Mazharimousavi:2014vza,Banados:2005hm} and higher
dimensional spacetimes\cite{Winstanley:2005fu,Martinez:2004nb,Kolyvaris:2009pc,
Gonzalez:2013aca,Nadalini:2007qi,
Lu:2014maa,Martinez:2006an,Martinez:2005di,Cardenas:2014kaa}.
Thermodynamically, AdS black holes with scalar hair are very interesting.
It was observed that static black
hole with minimally coupled scalar field hair
can transform to static BTZ black
hole in three dimensional AdS spacetimes\cite{Gegenberg:2003jr,Myung:2008ze}. 
A second order phase transition was disclosed to occur
between the four-dimensional AdS black hole with a minimally coupled
scalar field hair (Martinez-Troncoso-Zanelli(MTZ) black hole)
and the topological black hole with
hyperbolic horizon (TBH)\cite{Martinez:2004nb,Myung:2008ze}.
Furthermore, phase transitions among charged TBH, charged MTZ
black holes\cite{Martinez:2006an, Martinez:2010ti} and other exact 
hairy black hole solutions\cite{Gonzalez:2013aca} in the
four-dimensional AdS spacetimes were also
uncovered. It is of great interest to generalize the discussions
and find more examples on the relation between the dynamical
physical phenomenon and its corresponding thermodynamic properties.
In particular, the discussion for a rotating black hole in the STT is
still lack. In this work, we will concentrate our attention on the
rotating hairy AdS black holes in the three-dimensional
STT, which was recently
reconstructed in\cite{Xu:2014uha,Zou:2014gla}.

On the other hand, there exist two versions of STT:
one version is on the Einstein frame and the other is on the Jordan frame,
which can be related to the former by a conformal transformation.
One may have a nonminimally coupled scalar in
the Jordan frame, while one may have a minimally coupled scalar
in the Einstein frame. Here we want to ask whether the
intriguing thermodynamical relation between the rotating
hairy black holes and rotating BTZ black
holes in AdS spacetimes can be reflected in
dynamical properties so that it can have some
observational signatures to be detected in the both frames.
This is our main purpose here.
Considering that quasinormal modes of dynamical perturbations
are characteristic sounds of black holes, we expect
that the black hole phase transitions can be imprinted in the
dynamical perturbations in their surrounding geometries through
frequencies and damping times of the oscillations.
In the last few decades there have been extensive
research\cite{Gonzalez:2014tga,Rao:2007zzb,Koutsoumbas:2008pw,He:2010zb,
Liu:2011cu,Shen:2007xk,Koutsoumbas:2006xj,Liu:2014gvf},
where the relations between thermodynamical phase transitions and
dynamical perturbations have been discussed.
Besides, we will further discuss the static hairy black holes within both frames.

The paper is organized as follows: in the next section,
we review the thermodynamical properties of hairy black holes,
including the possible phase transitions between the
hairy black hole and BTZ black holes in the Jordan and Einstein frames.
In the third section, we will disclose that
this phase transition can be numerically
reflected by the QNM frequencies of
perturbations in both frames. We complete the paper with
conclusions and discussions in the fourth section.

\section{Thermodynamics of three dimensional AdS Black holes with scalar hair}

In this section, we consider the thermodynamics of hairy black holes
in the Einstein and Jordan frames. The action in Einstein
frame is given by \cite{Xu:2014uha}
\begin{eqnarray}
I=\frac{1}{2}\int \mathrm{d}^{3}x\sqrt{-g}\left(
R- \nabla_{\mu} \phi\nabla^{\mu} \phi -2V(\phi)\right),\label{minaction}
\end{eqnarray}
where $\phi(r)$ is the scalar field and the scalar potential $V(\phi)$
takes the form
\begin{eqnarray}
    V(\phi)&=-\frac{1}{l^2}\cosh^6\left(\frac{1}{2\sqrt{2}}\phi\right)
    +\frac{1}{l^2}\left(1+\mu l^2\right)
    \sinh^6\left(\frac{1}{2\sqrt{2}}\phi\right)\nonumber\\
    &-\frac{\alpha^2}{64}\sinh^{10}\left(\frac{1}{2\sqrt{2}}\phi\right)
    \cosh^{6}\left(\frac{1}
    {2\sqrt{2}}\phi\right)\bigg[\tanh^6\left(\frac{1}{2\sqrt{2}}\phi\right)
    \nonumber\\
    &-5\tanh^4\left(\frac{1}{2\sqrt{2}}\phi\right)
    +10\tanh^2\left(\frac{1}{2\sqrt{2}}\phi\right)-9\bigg]
\end{eqnarray}
with the parameters $\mu$, $\alpha$, and $l$ related to the negative cosmological
constant $\Lambda=-\frac{1}{l^2}$. Using a particular conformal factor of the form
$\check{g}_{\mu\nu}=\Omega^2g_{\mu\nu}$ and $\Omega=\cosh^2(\sqrt{8}\phi)$,
and defining a new scalar field $\check{\phi}(r)=\sqrt{8}\tanh(\sqrt{8}\phi)$
and scalar potential $U(\check{\phi})=V(\phi)\Omega^{-3}$,
the action above can be transformed into the corresponding
action in the Jordan frame\cite{Zou:2014gla}
\begin{eqnarray}
{\cal I}&=&\frac{1}{2}\int d^3x\sqrt{-\check{g}}(\check{R}-2\Lambda
-\check{g}^{\mu\nu}\nabla_{\mu}\check{\phi}\nabla_{\nu}\check{\phi}\nonumber\\
&&-\frac{1}{8}\check{R}\check{\phi}^2-2U(\check{\phi})) \label{i}
\end{eqnarray}
with
\begin{eqnarray}
U(\check{\phi})&=&\frac{1}{512}\left(\frac{1}{l^2}+\mu\right)\check{\phi}^6\nonumber\\
&&+\frac{\alpha^2\left(\check{\phi}^6-40\check{\phi}^4+640\check{\phi}^2
-4608\right)\check{\phi}^{10}}{512(\check{\phi}^2-8)^5}.\label{Vn}
\end{eqnarray}

\subsection{In the Jordan frame}

The action (\ref{i}) above admits the following solution \cite{Zou:2014gla}
\begin{eqnarray}
&&ds^2=-f(r)dt^2+f(r)^{-1}dr^2+r^2\left(d\varphi+\omega(r)dt\right)^2,\label{met}\\
&&f(r)=\mu B^2\left(3+\frac{2B}{r}\right)+\frac{\left(3r
+2B\right)^2\alpha^2B^4}{r^4}+\frac{r^2}{l^2},\nonumber\\
&&\omega(r)=-\frac{\alpha B^2\left(3r+2B\right)}{r^3}.
\end{eqnarray}
Here the parameter $B$ is related to the scalar field
\begin{eqnarray}
\check{\phi}(r)=\pm\sqrt{\frac{8B}{r+B}}.\label{phi}
\end{eqnarray}

For this rotating hairy black hole, the mass and
angular momentum can be calculated by adopting
the Brown-York method \cite{Brown:1992br}. The
quasilocal mass $m(r)$ at  $r$ takes the form
\cite{Brown:1994gs,Creighton:1995au,Chan:1995wj}
\begin{eqnarray}
m(r)=\sqrt{f(r)}E(r)-j(r)\omega(r),\label{3b}
\end{eqnarray}
where $E(r)=2(\sqrt{f_0(r)}-\sqrt{f(r)})$ is the
quasilocal energy at $r$ and
$j(r)=\frac{d\omega(r)}{dr}r^3$ is the quasilocal
angular momentum. Here, $f_0(r)=\frac{r^2}{l^2}$ is a background metric function
that determines the zero of the energy. As a result, the mass and
angular momentum of the black hole can be
obtained as
\begin{eqnarray}
M\equiv \lim_{r\rightarrow\infty}m(r)=-3\mu B^2,\quad
J\equiv \lim_{r\rightarrow\infty}j(r)=6\alpha B^2.\label{4b}
\end{eqnarray}
Thus the black hole metric coefficient
[Eq.~(\ref{met})] can be rewritten as
\begin{eqnarray}
&&f(r)=-M\left(1+\frac{2B}{3r}\right)+\frac{r^2}{l^2}+\frac{\left(3r+2B\right)^2J^2}{36r^4}.\label{nf}
\end{eqnarray}
The condition $-\mu\geq\frac{2\alpha}{l}$ is
needed to protect the cosmic censorship, which
puts the constraint $\frac{M}{J}\geq\frac{1}{l}$
for this black hole.

From $f(r_+)=0$, the horizon radius $r_+$ can be
expressed as $r_+=B\times \theta$, where $\theta$ is related to
the parameters $\mu$, $\alpha$ and $l$\cite{Zou:2014gla}.
Below, we set $l=1$ for simplicity.
Then the mass, temperature, heat capacity, angular momentum and entropy
of the rotating hairy black hole are obtained as
\begin{eqnarray}
&&M=\frac{J^2\left(2+3\theta\right)^2+36r_+^4\theta^2}{12r_+^2\theta\left(2+3\theta\right)},\nonumber\\
&&T=\frac{(1+\theta)\left[36r_+^4\theta^2-J^2(2
+3\theta)^2\right]}{24\pi r_+^3\theta^2\left(2+3\theta\right)},\nonumber\\
&&C=\frac{32\pi^2 r_+^4\theta\left(2+3\theta\right)T}
{(1+\theta)^2\left[12\theta^2 r_+^4+J^2\left(2+3\theta\right)^2\right]}\nonumber\\
&&J=Ma,\quad a=-\frac{2\alpha}{\mu},\quad S=\frac{4\pi\theta r_+}{1+\theta}.\label{TF}
\end{eqnarray}

In addition, by introducing a family of locally
nonrotating observers, the angular velocity for
these observers that move on orbits with constant $r$ is given by
\cite{Aliev:2007qi,Yue:2011et}
\begin{eqnarray}
\Omega=-\frac{g_{t\psi}}{g_{\psi\psi}}=-\omega(r)=\frac{\left(3r+2B\right)J}{6r^3}.\label{14b}
\end{eqnarray}
When approaching the black hole horizon, the
angular velocity $\Omega_+$ turns out to be
\begin{eqnarray}
\Omega_+&=&-\omega(r_+)=\frac{\left(3r_{+}+2B\right)J}{6r_+^3}\nonumber\\
&=&\frac{(2+3\theta)J}{6\theta r_+^2}.\label{oh}
\end{eqnarray}
Combining these quantities, $M$, $T$, and $\Omega_+$, we
can verify that the first law of thermodynamics
holds in this case
\begin{eqnarray}
dM=TdS+\Omega_+ dJ
\end{eqnarray}
and the Smarr relation can be found
\begin{eqnarray}
M-\Omega_{+} J=\frac{1}{2}TS.
\end{eqnarray}

\begin{figure*}
  \includegraphics{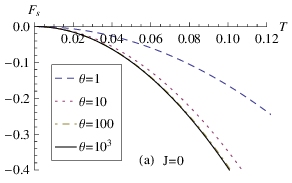}
   \hfill%
  \includegraphics{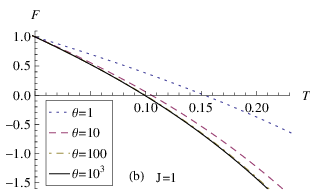}
   \hfill%
  \includegraphics{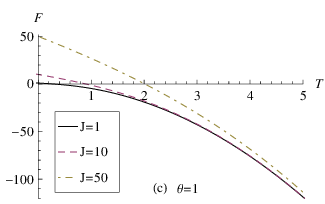}
  \label{f.first}
  \caption{ The free energies $F$ of hairy black hole versus
  the temperature $T$ in the Jordan frame. }
     \label{fig.1}
\end{figure*}

In general, thermodynamical stabilities can be disclosed by examining the free energy.
It has the following expression in the canonical ensemble (fixed $J$)
\begin{eqnarray}
F_r=M-TS=\frac{J^2\left(2+3\theta\right)^2-12r_+^4\theta^2}{4r_+^2\theta\left(2+3\theta\right)}.\label{FF}
\end{eqnarray}
Here the subscript $``r"$ denotes the rotating black hole.
Firstly, we consider the static case $(J=0)$. Using the relation (\ref{TF}),
the free energy of the static hairy black hole can be written as
\begin{eqnarray}
F_s=-\frac{\theta(2+3\theta)}{3(1+\theta)^2}(2\pi T)^2=-\xi(\theta)(2\pi T)^2.\label{Fs}
\end{eqnarray}
Here the subscript $``s"$ denotes the static black hole.
It is easily verified that $\xi(\theta)$ is a monotone
increasing function of $\theta$ because of $\frac{d\xi(\theta)}{d\theta}>0$.
Therefore the free energy $F_s$ becomes more negative
with the increasing of $\theta$, so that this static hairy
black hole becomes more stable, see fig.~\ref{fig.1}(a).
For a rotating hairy black hole, we can rewrite eqs.~(\ref{TF})(\ref{FF}) as
\begin{eqnarray}
T=\frac{F_r\sqrt{F_r^2+3J^2}-F_r^2-J^2}{\sqrt{2\xi(\theta)}(\sqrt{F_r^2
+3J^2}-F_r)^{3/2}}\equiv\frac{\Upsilon(F_r,J)}{\sqrt{\xi(\theta)}}.\label{ST}
\end{eqnarray}
Consider the function $\Upsilon(F_r,J)$, we obtain
\begin{eqnarray}
&\frac{\partial\Upsilon(F_r,J)}{\partial F_r}=\frac{2F_r\sqrt{F_r^2+3J^2}
-\left(2F_r^2+3J^2\right)}{2\sqrt{2}\pi\left(\sqrt{F_r^2+3J^2}-F_r\right)^{5/2}}<0,\\
&\frac{\partial\Upsilon(F_r,J)}{\partial J}=\frac{J\left(\sqrt{F_r^2+3J^2}
-F_r\right)}{2\sqrt{2}\pi\left(\sqrt{F_r^2+3J^2}-F_r\right)^{5/2}}>0,
\end{eqnarray}
which imply that $\Upsilon(F_r,J)$ is a monotone increasing function of $J$
or a monotone decreasing function of $F_r$. Consequently,
the free energy $F_r$ with any fixed temperature $T$ will reach smaller
values when taking
smaller values of $J$ or larger values of $\theta$
because of the monotone increasing function $\xi(\theta)$ of $\theta$, or vice versa.
These properties are shown in fig.~\ref{fig.1}(b)(c).
In addition, the function $\xi(\theta)$ increases more quickly
than $\Upsilon(F_r,J)$ of $J$ on account
of $\frac{d^2\Upsilon(F_r,J)}{dJ^2}<0$ and $\frac{d^2\xi(\theta)}{d\theta^2}>0$.

Another solution of this theory is the rotating BTZ black hole,
which is obtained from the action (\ref{i}) provided
that $\check{\phi}(r)=0$. The corresponding metric is
\begin{eqnarray}
&ds^2=-f(\rho)dt^2+f(\rho)^{-1}d\rho^2+\rho^2\left(d\varphi+\Omega(\rho)dt\right)^2,\label{Bmet}\\
&f(\rho)=-\hat{M}+\rho^2+\frac{\hat{J}^2}{4\rho^2},\qquad \qquad
\Omega(\rho)=-\frac{\hat{J}^2}{2\rho^2}.\label{BS}
\end{eqnarray}
The thermodynamic quantities in the canonical ensemble(fixed $J$),
such as the temperature, mass, heat capacity, entropy, angular velocity
and free energy of the rotating BTZ black hole are given by
\begin{eqnarray}
&&\hat{M}=\rho_+^2+\frac{\hat{J}^2}{4\rho_+^2}
\quad,\hat{\Omega}_+=\frac{\hat{J}}{2\rho_+^2}, \quad
\hat{S}=4\pi\rho_+,\nonumber\\
&&\hat{T}=\frac{4\rho_+^4-\hat{J}^2}{8\pi \rho_+^3},\quad
\hat{C}=\frac{32\pi^2\rho_+^4 T}{4\rho_+^4+3\hat{J}^2},\label{BTZ}\\
&&\hat{F}_r=\frac{3\hat{J}^2-4\rho_+^4}{4\rho_+^2}.\label{BTZF}
\end{eqnarray}
We can also rewrite eq.~(\ref{BTZF}) into
\begin{eqnarray}
\hat{T}=\frac{\hat{F}_r\sqrt{\hat{F}_r^2+3\hat{J}^2}
-\hat{F}_r^2-\hat{J}^2}{\sqrt{2}(\sqrt{\hat{F}_r^2
+3\hat{J}^2}-\hat{F}_r)^{3/2}}=\hat{\Upsilon}(\hat{F}_r,\hat{J}).\label{BTZT}
\end{eqnarray}
Clearly, the function $\hat{\Upsilon}(\hat{F}_r,\hat{J})$ shares
the same form as the function $\Upsilon(F_r,J)$,
whose function characteristic has been discussed above.

Notice that the specific heat (eq.~(\ref{TF})) of the hairy black hole
is always positive, which means that the scalar black hole can
always reach thermal equilibrium with a heat bath. Moreover,
a BTZ black hole with a vanishing scalar field can also be
at equilibrium with the heat bath. This
raises the question of whether the hairy black hole could decay into the BTZ
black hole. In order to find possible phase transition between the hairy and
undressed states, we introduce the matching for the temperature
$T=\hat{T}$ and angular momentum $J=\hat{J}$, and obtain
\begin{eqnarray}
\hat{\Upsilon}(\hat{F}_r,J)=\frac{\Upsilon(F_r,J)}{\sqrt{\xi(\theta)}}\label{10a}.
\end{eqnarray}
Taking into account $0<\sqrt{\xi(\theta)}<1$, we can obtain
\begin{eqnarray}
\hat{\Upsilon}(\hat{F}_r,J)>\Upsilon(F_r,J)\Rightarrow \hat{F}_r<F_r.\label{10b}
\end{eqnarray}
This indicates that, with same temperature $T$ and angular momentum $J$,
the rotating BTZ black hole always has smaller
free energy which is a thermodynamically more preferred phase compared to
the rotating hairy black hole. More importantly, this conclusion is universal
and irrelevant to any specific positive values of $J$ and $\theta$.
Thus, there exists a possible
phase transition for this rotating hairy black hole to become rotating BTZ black hole,
provided that there are some thermal fluctuations.
In static case, we can directly calculate
the difference of free energies from eqs.~(\ref{Fs})(\ref{BTZF})
\begin{eqnarray}
\Delta F_s=F_s-\hat{F}_s=\frac{3+4\theta}{3(1+\theta)^2}(2\pi T)^2.
\end{eqnarray}
Since the constant $\theta$ always remains
positive\cite{Zou:2014gla},
we have $\Delta F_s>0$, namely $F_s>\hat{F}_s$.
So, the static BTZ black hole is more thermodynamically preferred.
When $\theta\rightarrow +\infty$, $\Delta F_s$ approaches zero, then the free
energies of two black holes become consistent.

\subsection{In the Einstein frame}

Now we generalize the above discussions to the three-dimensional
hairy black holes in the Einstein frame. As shown in \cite{Xu:2014uha},
the action (\ref{minaction}) admits the following solution
\begin{eqnarray}
&\mathrm{d}s^{2}=-f_1(H)\mathrm{d}t^{2}+\frac{\mathrm{d}r^{2}}{f_2(H)}
+r^{2}\bigg(\mathrm{d}\psi+\omega(H)\mathrm{d}t\bigg)^{2},\nonumber\\
&f_1(H)=\frac{H^2f(H)}{(H+B)^2},\quad
f_2(H)=\frac{(H+2B)^2f(H)}{(H+B)^2}\label{minmetric},
\end{eqnarray}
where the function $f(H)$ and $\omega(H)$ read as
\begin{eqnarray}
&f(H)=3\mu B^2+\frac{2\mu B^3}{H}+\frac{\alpha^2 B^4(3H+2B)^2}{H^4}
+\frac{H^2}{l^2},\nonumber\\
&\omega(H)=-\frac{\alpha B^2(3H+2B)}{H^3}.
\end{eqnarray}
Here the parameter $B$ is related to the scalar field
\begin{eqnarray}
&&\phi(r)=2\sqrt{2}arctanh\sqrt{\frac{B}{H(r)+B}},\nonumber\\
&&H(r)=\frac{1}{2}\left(r+\sqrt{r^2+4Br}\right) \label{ph}.
\end{eqnarray}

From $f(H)=0$, the horizon radius $H_+$ can be expressed as $H_+=B\times h$,
where $h$ is related to the parameters $\mu$, $\alpha$ and $l$ \cite{Xu:2014uha}.
Moreover, the corresponding thermodynamical quantities of the rotating
hairy black hole are obtained as
\begin{eqnarray}
M&=&\,{\frac {{J}^{2}(3h+2)^2+36h^2\,H_+^{4}}{12h \left( 3h+2\right)H_+^{2} }},
\quad \Omega_+=\frac{(3h+2)J}{6h H_+^2},\nonumber\\
T&=&\frac{(1+h)\left[36H_+^4h^2-J^2(2+3h)^2\right]}{24\pi h^2\left(2+3h\right) H_+^3},
\quad S=\frac{4\pi h H_+}{(h+1)},\nonumber\\
C&=&\frac{32\pi^2 H_+^4h\left(2+3h\right)T}
{(1+h)^2[12h^2 H_+^4+J^2\left(2+3h\right)^2]},\nonumber\\
F_r&=&\,{\frac {\,{J}^{2}(3h+2)^2-12h^2\,H_+^{4}}{4h \left(3h+2\right)H_+^{2}}}.\label{minthermo}
\end{eqnarray}
Notice that these expressions are similar to their counterparts
(\ref{TF})(\ref{oh})(\ref{FF}) for the hairy
black hole in the Jordan frame disclosed in fig.~\ref{fig.1}.
This implies that the static/rotating hairy black hole in the
Einstein frame possesses the same thermodynamical properties,
compared with the static/rotating hairy black
holes in the Jordan frame. Moreover, there also exists a nonvanishing
possibility for a static/rotating hairy black hole to decay
into a static/rotating BTZ black hole.
In other words, we find that the black hole
thermodynamical stabilities in both frames are equivalent.

\section{QNMs of electromagnetic field perturbations}

\subsection{In the Jordan frame}

To reflect the thermodynamical stabilities in dynamical perturbations,
we calculate the QNMs of electromagnetic field perturbations around
this hairy black hole. The electromagnetic perturbations are governed
by Maxwell's equations
\begin{eqnarray}
F^{\mu\nu}_{~~~;\nu}=0, \quad
F_{\mu\nu}=\partial_{\mu}A_{\nu}-\partial_{\nu}A_{\mu}.\label{el}
\end{eqnarray}
As the background is circularly symmetric, it would be advisable to expand $A_{\mu}$
in 3-dimensional vector spherical harmonics
\begin{eqnarray}
A_{\mu}(t,r,\varphi)=
\left(
\begin{array}{c}
 P(r)  \\
 S(r) \\
 Q(r)
\end{array}
\right)e^{-i\omega t+i m \varphi},\label{al}
\end{eqnarray}
where $m$ is our angular quantum number and $\omega$ is the frequency.
With the ansatz (\ref{met}), we substitute the expansion (\ref{al}) into Maxwell's
equations (\ref{el}), and obtain
\begin{eqnarray}
F^{\varphi\nu}_{~~;\nu}&=&\frac{d}{dr}\left(rN(r)\omega(r)^2+rK(r)\omega(r)\right)\nonumber\\
&&-\frac{d}{dr}\left(\frac{f(r)N(r)}{r}\right)
+\frac{\omega r\left(mP(r)+\omega Q(r)\right)}{r^2f(r)}=0,\nonumber\\
F^{r\nu}_{~~;\nu}&=&-mr^2N(r)\omega(r)+\left(mf(r)-\omega r^2\omega(r)\right)N(r)\nonumber\\
&&-\left(\omega+m\omega(r)\right)r^2K(r)=0,\nonumber\\
F^{t\nu}_{~~;\nu}&=&\frac{d}{dr}\left(rN(r)\omega(r)+rK(r)\right)\nonumber\\
&&-\frac{m\left(m P(r)+\omega Q(r)\right)}{rf(r)}=0,\label{se3}
\end{eqnarray}
where $N(r)=imS(r)-\frac{dQ(r)}{dr}$ and $K(r)=i\omega S(r)+\frac{dP(r)}{dr}$.

Based on these equations, a second order differential radial equation for the electromagnetic
perturbation can be derived as
\footnote{For a massless scalar field perturbation around the three dimensional hairy
black hole in the Jordan frame, the Klein-Gordon equation is given by
\begin{eqnarray}
\frac{1}{\sqrt{-g}}\partial_\mu\left(g^{\mu\nu}\sqrt{-g}\partial_\nu\right)\Psi=0.\nonumber
\end{eqnarray}
Introducing the following ansatz for the field
\begin{eqnarray}
\Psi=\frac{\psi(r)}{r^{1/2}}e^{-i\omega t+i m \varphi},\nonumber
\end{eqnarray}
the radial equation for scalar field perturbation takes the same form as eq.~(\ref{sre}).}
\begin{eqnarray}
\psi''(r)&+&\frac{f'(r)\psi'(r)}{f(r)}+\left[\frac{(m\omega(r)
+\omega)^2}{f^2(r)}+\frac{1}{4r^2}\right.\nonumber\\
&&\left.-\frac{m^2}{r^2f(r)}-\frac{f'(r)}{2rf(r)}\right]\psi(r)=0,\label{sre}
\end{eqnarray}
where the wavefunction $\psi(r)$ is set to be a combination of
functions $f(r)$, $\omega(r)$ and $N(r)$
\begin{eqnarray}
\psi(r)=\frac{f(r)}{r^{1/2}}\frac{\omega N(r)}{m\omega(r)+\omega}.
\end{eqnarray}

\begin{figure}
  \includegraphics{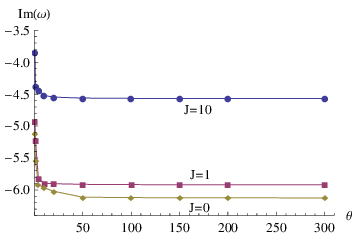}
  \caption{Lowest QNM frequencies of electromagnetic perturbations with
  $m=1$ around hairy black holes at $T=0.5$ in the Jordan frame.}
   \label{fig.2}
\end{figure}

Defining the tortoise coordinate $dr_*=dr/f(r)$, the radial
wave equation can be expressed as
\begin{eqnarray}
\partial^2_{r_*}\psi(r)+V(r)\psi(r)=0.
\end{eqnarray}
where the generalized potential $V(r)$ is given by
\begin{eqnarray}
V(r)=(m\omega(r)+\omega)^2+\frac{f(r)^2}{4r^2}
-\frac{m^2f(r)}{r^2}-\frac{f'(r)f(r)}{2r}.\nonumber
\end{eqnarray}
At the AdS boundary $r\rightarrow +\infty$, the generalized potential $V(r)$
diverges. Then, we need $\psi(r)=0$\cite{Cardoso:2004fi}.
From eq.~(\ref{sre}), the incoming wavefunction $\psi(r)$ near the horizon reads as
\begin{eqnarray}
\psi(r)\sim (r-r_+)^{-\frac{i\kappa}{4\pi T}}, \quad
\kappa=\omega+m\omega(r_+)\label{R1}.
\end{eqnarray}
Defining $\psi(r)$ as $\varphi(r)exp[-i\int\frac{\kappa}{f(r)}dr]$,
where $exp[-i\int\frac{\kappa}{f(r)}dr]$ asymptotically approaches to the ingoing wave
near horizon, eq.~(\ref{sre}) becomes
\begin{eqnarray}
&&4r^2f(r)\varphi''(r)+4r^2\varphi'(r)\left[f'(r)-2i\kappa\right]
+\left[f(r)-2(2m^2\right.\nonumber\\
&&\left.+r f'(r))+\frac{4r^2((m\omega(r)+\omega)^2
-\kappa^2)}{f(r)}\right]\varphi(r)=0\label{KG}
\end{eqnarray}
so that $\varphi(r_+)=1$ in case of $r\rightarrow r_+$.
Then we can numerically solve eq.~(\ref{KG}) and find the QNM frequencies
under the boundary conditions by adopting the shooting method.
The lowest QNM frequencies of perturbations with $m=1$
around this static and rotating hairy black holes at $T=0.5$ are shown in fig.~\ref{fig.2}.
Fixing the angular momentum, the imaginary parts of both QNM
frequencies become more negative with the increase of $\theta$, and then almost
remains unchanged when $\theta$ is big enough. Moreover, the rotating
hairy black hole with smaller values of $J$ is more
stable, since the imaginary parts of QNM frequencies are more negative. These
results exactly reflect the thermodynamic stabilities of
the static and rotating hairy black holes disclosed in fig.~1.

\begin{figure*}
\includegraphics{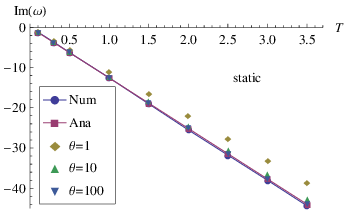}
\hfill%
\includegraphics{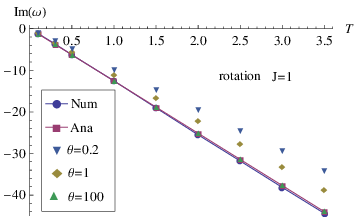}
\caption{Lowest QNM frequencies of electromagnetic perturbations
 with $m=1$ for hairy black holes at $T=0.5$ in the Jordan frame.}
\label{fig.3}
\end{figure*}

On the other hand, the QNM frequencies for BTZ and hairy black holes with $m=1$
are shown in fig.~\ref{fig.3}. One can see that the numerical results of QNM frequencies
agree excellently with analytical ones \cite{Birmingham:2001hc}
in (rotating and static) BTZ black holes. Moreover, with the growth of $\theta$,
the imaginary parts for a hairy black hole become more
negative, and then approach the curves of a BTZ back hole.
These results match the thermodynamical stability analysis above.

\subsection{In the Einstein frame}

\begin{figure}
\onefigure{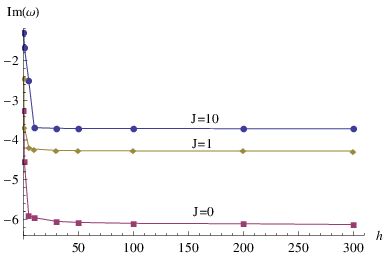}
\caption{Lowest QNM frequencies of electromagnetic perturbations
with $m=1$ around hairy black holes at $T=0.5$ in the Einstein frame.}
\label{fig.4}
\end{figure}

\begin{figure*}
\includegraphics{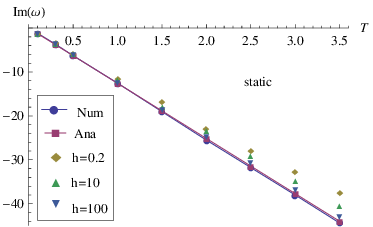}
\hfill%
\includegraphics{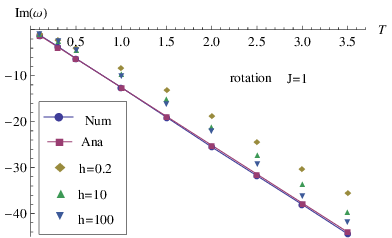}
\caption{Lowest QNM frequencies of electromagnetic perturbations
 with $m=1$ for hairy black holes at $T=0.5$ in the Einstein frame.}
\label{fig.5}
\end{figure*}

Now we calculate the QNMs of electromagnetic field perturbations around
this rotating/static hairy black hole in the Einstein frame.
The differential equation for electromagnetic perturbation 
in the Einstein frame can be obtained as
\footnote{For a massless scalar field perturbation around the three dimensional hairy
black hole in the Einstein frame, it will be found that
the perturbation equation takes the same form as eq.~(\ref{minKG3}).}
\begin{eqnarray}
&&\varphi''(H)+\varphi'(H)\left[\frac{f'(H)}{f(H)}-2i\kappa\zeta\right]
+\left[\frac{4B+H}{4H(B+H)^2}\right.\nonumber\\
&&\left.-\frac{m^2}{f(H)}-\frac{(B+H)f'(H)}{2H^2\zeta f(H)}+\frac{\left(m\omega(H)
+\omega\right)^2}{f(H)^2}\right.\nonumber\\
&&\left.-\frac{\kappa^2\zeta^2}{f(H)^2}
+\frac{2iB^2\zeta^2\kappa}{(B+H)^3f(H)}\right]\varphi(H)=0\label{minKG3}
\end{eqnarray}
with
\begin{eqnarray}
\zeta=\frac{(B+H)^2}{H(2B+H)},\quad
\kappa=\frac{(2+h)h}{(1+h)^2}\left(\omega+\omega(H_+\right).\nonumber
\end{eqnarray}
Here function $\varphi(H)$ equals to 1 in case of $H\rightarrow H_+$, and
vanishes at AdS infinity.
By calculating numerically eq.~(\ref{minKG3}) under
the boundary conditions, we find that, in the Einstein frame, the behaviors
of lowest QNM frequencies are consistent with the thermodynamic stabilities
of these hairy black holes disclosed in figs.~\ref{fig.4},\ref{fig.5}.

\section{Conclusions and Discussions}

By calculating the QNMs of electromagnetic
perturbations, we found that the rotating hairy
black holes with larger values of $\theta$ (in the Jordan frame)
or $h$ (in the Einstein frame), and smaller values of $J$
can be more dynamically stable objects in the canonical ensemble. The
disclosed dynamical stability properties are
consistent with the thermodynamical stability for these hairy black holes.
On the other hand, compared with the rotating hairy black hole,
the rotating BTZ black hole is dynamically more stable in both frames,
due to the fact that it has larger absolute imaginary QNM frequencies.
This property is again in agreement with the fact that
the rotating BTZ black hole shares lower free energy 
and is more thermodynamically preferred. Similar observations
on the dynamical phenomenon and its relation to
thermodynamics have been disclosed between
the static hairy black hole and
its BTZ black hole counterpart in both frames.
Considering that the QNM is expected
to be detected and has strong astrophysical
interest, we hope that the dynamical probe can
really present us the observational signature of
the thermodynamic property of black holes.

Recently, Yun Soo Myung et al.\cite{Myung:2014nua}
found that the black hole (in)stability is independent
of the frame which shows that the two frames are equivalent to each other.
Similar equivalence of the Einstein and Jordan frames also appeared in the
Higgs inflation\cite{Postma:2014vaa}. However,
S.Bahamonde et al.\cite{Bahamonde:2016wmz}
found the frame dependence of $F(R)$ gravity singularities in Jordan and Einstein Frames.
Moreover, Faraoni et al.\cite{Faraoni:2006fx,Banerjee:2016lco} showed that,
with some proper interpretation,
the two versions are actually equivalent
at the classical level, while inequivalent at the quantum level.
Here our present work provides a example to support the claim of the equivalence
of two frames by means of the hairy black hole thermodynamics
and corresponding dynamical properties in both frames.
However, whether the two versions of STT are equivalent or not
in the classical gravity is still generates plenty of controversy.
This deserves research in further.

\acknowledgments
This work is supported by Natural Science Foundation
of Higher Institutions of Jiangsu Province (Grant No.16KJB140020)
and Natural Science Foundation for Youths of Jiangsu Province (Grant No.BK20160452).

\end{document}